\documentclass[twocolumn]{autart}    

\usepackage{graphicx}          
\usepackage[dvips]{epsfig}    

\usepackage{amssymb,amsmath,amsfonts,subfigure,enumerate,natbib}
\usepackage{dsfont}

\def\be{\begin{equation}}
\def\ee{\end{equation}}
\def\ben{\begin{equation*}}
\def\een{\end{equation*}}

\newcommand{\epf}{$\Box$}
\newcommand{\dfb}{\stackrel{\Delta}{=}}

\def\re{\mathop{Re}\nolimits}
\def\im{\mathop{Im}\nolimits}
\def\r{\mathbb R}
\def\ones{\mathds{1}}
\def\C{\mathbb C}

\begin{document}

\begin{frontmatter}

\title{Differential Inequalities in Multi-Agent Coordination and Opinion Dynamics Modeling.
\thanksref{footnoteinfo}}

\thanks[footnoteinfo]{
Partially supported by the European Research Council under grant ERC-StG-307207, NWO under grant vidi-438730
and the Russian Foundation for Basic Research (RFBR) under grants 17-08-01728, 17-08-00715 and 17-08-01266. Theorem~\ref{thm.1} is supported solely by the Russian Science
Foundation grant 14-29-00142. Theorem~\ref{thm.aggreg} is obtained supported solely by
the Russian Science Foundation grant 16-19-00057.}

\author[Delft,IPME,ITMO]{Anton V. Proskurnikov}\ead{anton.p.1982@ieee.org}\quad
\author[Gro]{Ming Cao}\ead{m.cao@rug.nl}

\address[Delft]{Delft Center for Systems and Control, Delft University of Technology, The Netherlands}
\address[IPME]{Institute for Problems of Mechanical Engineering, Russian Academy of Sciences, St. Petersburg, Russia}
\address[ITMO]{Chair of Mathematical Physics and Information Theory, ITMO University, St.-Petersburg, Russia}
\address[Gro]{Engineering and Technology Institute (ENTEG), University of Groningen, The Netherlands}
\begin{keyword}
Multi-agent systems, cooperative control, distributed algorithm, complex network
\end{keyword}

\begin{abstract}                          

Many distributed algorithms for multi-agent coordination employ the simple averaging dynamics, referred to as the \emph{Laplacian flow}.
Besides the standard consensus protocols, examples include, but are not limited to, algorithms for aggregation and containment control, target surrounding, distributed optimization and
models of opinion formation in social groups. In spite of their similarities, each of these algorithms has been studied using separate mathematical techniques.
In this paper, we show that stability and convergence of many coordination algorithms involving the Laplacian flow dynamics follow from the general \emph{consensus dichotomy} property of a special differential inequality. The consensus dichotomy implies that any solution to the differential inequality is either unbounded or converges to a consensus equilibrium. In this paper, we establish the
dichotomy criteria for differential inequalities and illustrate their applications to multi-agent coordination and opinion dynamics modeling.
\end{abstract}

\end{frontmatter}

\section{Introduction}

Distributed algorithms for multi-agent coordination have various applications to science and engineering, including control of robotic formations, scheduling of sensor networks, optimization and filtering, modeling biological and social systems. The relevant results are discussed in the works~\citep{RenBeardBook,MesbahiEgerBook,RenCaoBook,SavkinBook2015,ProCao16-EEEE,BulloBook-Online,ProTempo:2017-1} and references therein. A ``benchmark'' problem in multi-agent control is to establish consensus (that is, agreement on some quantity of interest) among the agents interacting over a general graph. A simple consensus algorithm, originated from some opinion formation models~\citep{ProTempo:2017-1}, is called the \emph{Laplacian flow}~\citep{BulloBook-Online}.
Being a counterpart~\citep{Ferrari:06} of the well-known heat equation, which is used in physics to describe diffusion processes, this algorithm employs the Laplacian matrix $L(t)$ of the interaction graph
\be\label{eq.proto}
\dot x(t)=-L(t)x(t).
\ee
The state vector's $i$th component $x_i(t)$ stands for some value, owned by agent $i$ and representing some quantity of interest (e.g. temperature or altitude).
The Laplacian flow dynamics~\eqref{eq.proto} describe the agents' interactions in order to agree on this quantity, which means that all $x_i(t)$ converge to a common value.
Numerous extensions of the protocol~\eqref{eq.proto} have been studied in the literature~\citep{RenBeardBook,RenCaoBook,CaoYuRenChen:2013}.

The effect of the interaction graph on establishing consensus has been studied up to a certain exhaustiveness by using the results on convergence of products of stochastic matrices~\citep{CaoMorse:08Part1,RenBeardBook} and special Lyapunov functions~\citep{Moro:04,LinFrancis:07,Muenz:11}. Consensus is established under rather mild assumption of ``repeated'' (``uniform'') connectivity of the graph; this condition can be further relaxed for some special types of graphs~\citep{TsiTsi:13}. The algorithms~\eqref{eq.proto} have inspired numerous protocols for synchronization of general dynamical systems~\citep{RenCaoBook,CaoYuRenChen:2013}.

In spite of the progress in the analysis of consensus algorithms, the relevant mathematical techniques are not directly applicable to other distributed coordination algorithms,
employing the idea of the Laplacian flows. The algorithms for containment and aggregation control~\citep{RenCaoBook,ShiHong:09}, target surrounding~\citep{LouHong:15} and convex optimization~\citep{ShiJohanssonHong:13}, as well as some models of opinion dynamics~\citep{Altafini:2013} are similar in spirit to consensus protocols; however,
each of the mentioned algorithms has been examined by using separate mathematical techniques. It appears, however, that the mentioned algorithms can be analyzed in a unified way, since they reduce to the following differential \emph{inequalities}, associated to the Laplacian flow dynamics~\eqref{eq.proto}
\be\label{eq.diff}
\dot x(t)\le -L(t)x(t).
\ee

The one-sided inequalities~\eqref{eq.diff} may seem very ``loose'' restrictions on the solutions $x(t)$. Nevertheless, under natural connectivity assumptions any solution,
which is \emph{semi-bounded} from below, converges to a consensus equilibrium.
In particular, the solutions of the differential inequality split into two groups: unbounded solutions and converging ones.
For ordinary differential equations the corresponding property is often referred to as the equation's \emph{dichotomy}~\citep{Yak:88SCL}.
In this paper, we establish the dichotomy properties of the differential inequalities~\eqref{eq.diff} and demonstrate their applications to the problems of multi-agent coordination, distributed optimization algorithms and some models of opinion formation. Some results have been reported in the conference paper~\citep{ProCao16-4}.

The paper is organized as follows. Section~\ref{sec.prelim} introduces some preliminary concepts and notation.
Section~\ref{sec.ineq1} introduces the Laplacian differential inequalities and presents their dichotomy conditions. Section~\ref{sec.example} illustrates applications of the main results, whose proofs are given in Section~\ref{sec.proof}. Section~\ref{sec.concl} concludes the paper.

\section{Preliminaries and notation}\label{sec.prelim}

We use $\ones_N$ to denote the column vector of ones $\ones_N\dfb(1,1,\ldots,1)^{\top}\in\r^N$. For two vectors $x,y\in\r^N$
we write $x\le y$ (or $y\ge x$) if $x_i\le y_i\,\forall i$.
Given a vector $x\in\r^N$, $|x|\dfb\sqrt{x^{\top}x}$ denotes its Euclidean norm.
Given a complex number $z\in\C$, $z^*$ denotes its complex conjugate.

\begin{figure}
\center
\includegraphics[height=3cm]{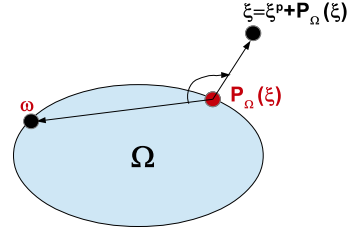}
\caption{The projection onto a closed convex set}\label{fig.proj}
\end{figure}

Given a closed convex set $\Omega\subset\r^d$, the \emph{projection} operator $P_{\Omega}:\xi\in\r^d\mapsto P_{\Omega}(\xi)\in\Omega$ is defined. Denoting $\xi^p\dfb \xi-P_{\Omega}(\xi)$, the distance from $\xi$ to $\Omega$ is given by $d_{\Omega}(\xi)\dfb|\xi^p|=\min_{\omega\in\Omega}|\xi-\omega|$.
For an arbitrary $\omega\in\Omega$, one has $|\xi-P_{\Omega}(\xi)-\alpha(\omega-\xi)|\ge |\xi^p|$ for any $\alpha\in [0;1]$,
entailing that $(\omega-P_{\Omega}(\xi))^{\top}\xi^p\le 0$, that is, $\measuredangle(\omega-P_{\Omega}(\xi),\xi-P_{\Omega}(\xi))\ge\pi/2$ (Fig.~\ref{fig.proj}). Therefore
\begin{gather}\label{eq.projection1}
(\omega-\xi)^{\top}\xi^p\le -|\xi^p|^2\quad\forall\xi\in\r^d,\omega\in\Omega\\
\begin{aligned}
(\xi_2-\xi_1)^{\top}\xi_1^p&=(\xi_2^p)^{\top}\xi_1^p+(P_{\Omega}(\xi_2)-\xi_1)^{\top}\xi_1^p\overset{\eqref{eq.projection1}}{\le}\\
&\le (\xi_2^p)^{\top}\xi_1^p-|\xi_1^p|^2\quad\forall\xi_1,\xi_2\in\r^d,
\end{aligned}\label{eq.projection}\\
(\xi_2-\xi_1)^{\top}(\xi_2^p-\xi_1^p)\overset{\eqref{eq.projection}}{\ge} |\xi_2^p-\xi_1^p|^2\quad\forall\xi_1,\xi_2\in\r^d\label{eq.non-exp}.
\end{gather}
The inequality~\eqref{eq.non-exp} implies that the mapping $\xi\mapsto\xi^p$ is non-expansive $|\xi_2-\xi_1|\ge |\xi_2^p-\xi_1^p|$. Furthermore, as shown in \citep[Lemma~2]{ShiHong:09}, the function $\xi\mapsto d_{\Omega}(\xi)^2=|\xi^p|^2$ is $C^1$-smooth with the gradient
\be\label{eq.gradient}
\nabla\left(d_{\Omega}(\xi)^2\right)=2\xi^p=2\left(\xi-P_{\Omega}(\xi)\right).
\ee
We assume that the reader is familiar with the standard concepts of graph theory, related to directed graphs, such as walks, strong connectivity and strongly connected components, see e.g.~\citep{HararyBook:1965,BulloBook-Online}. Henceforth each graph is directed and weighted, being thus a triple
$G=(V,E,A)$, where $V$ stands for the set of nodes, $E\subset V\times V$ is a set of arcs and $A=(a_{ij})_{i,j\in V}$ is an adjacency matrix: $a_{ij}>0$ if $(j,i)\in E$ and $a_{ij}=0$ otherwise.
By default, the adjacency matrix is assumed to be binary ($a_{ij}\in\{0,1\}$), such a graph is denoted simply by $G=(V,E)$. Any non-negative square matrix $A\in\r^{N\times N}$ can be associated to the graph $G[A]\dfb (V_N, E[A],A)$, where
$V_N\dfb\{1,\ldots,N\}$ and $E[A]\dfb\{(j,i):a_{ij}>0\}$. The \emph{Laplacian} matrix of this graph is defined as follows
\be\label{eq.lapl}
L[A]=(l_{ij})_{i,j=1}^N,\quad
l_{ij}\dfb
\begin{cases}
-a_{ij},\;\; i\ne j\\
\sum\limits_{j\ne i}a_{ij},i=j.
\end{cases}
\ee
A graph is \emph{quasi-strongly connected} (QSC) if a ``root'' node exist, from which all other nodes are reachable via walks, or, equivalently, the graph has a directed \emph{spanning tree}~\citep{RenBeardBook}. Given an adjacency matrix $A=(a_{ij})$ and $\delta>0$, define its ``truncation''  $A^{\delta}=(a_{ij}^{\delta})$ as follows: $a_{ij}^{\delta}=a_{ij}$ if $a_{ij}\geq\delta$ and otherwise $a_{ij}^{\delta}=0$. The graph $G[A]$ is strongly (quasi-strongly) \emph{$\delta$-connected} if its subgraph $G[A^{\delta}]$, obtained by removing ``light-weight'' arcs, is strongly (respectively, quasi-strongly) connected.

\section{Dichotomies of Differential Inequalities}\label{sec.ineq1}

The proofs of all theorems from this section can be found in Section~\ref{sec.proof}.
Henceforth we assume that a time-varying graph $G(t)=(V_N,E(t),A(t))$ without self-loops ($a_{ii}(t)=0\,\forall i$) is given, which corresponds to the Laplacian $L(t)$.
We are interested in the solutions of the \emph{differential inequality}~\eqref{eq.diff}.
The function $x:[0;\infty)\to\r^N$ is said to be a solution to the inequality~\eqref{eq.diff} if it is absolutely continuous and satisfies~\eqref{eq.diff} for almost any $t\ge 0$.

Throughout this section we also adopt the following assumption, which usually holds in practice and simplifies the further analysis; in some of the subsequent results it can be relaxed.
\begin{assum}\label{ass.l1-loc}
The functions $a_{ij}(t)$ are bounded.
\end{assum}

Under Assumption~\ref{ass.l1-loc}, for any $x(0)$ the solution to the equation~\eqref{eq.proto} exists that satisfies~\eqref{eq.diff} and is bounded. The inequality~\eqref{eq.diff} has also many \emph{unbounded} solutions, e.g. the function $x(t)\dfb x(0)-tc\ones_N$ satisfies~\eqref{eq.diff} for large $c>0$ since $L(t)$ is bounded and $L(t)\ones_N=0$.
At the same time, any solution to~\eqref{eq.diff} is \emph{upper-semibounded}.
\begin{lem}\label{lem.above}
For any solution $x(t)$ of~\eqref{eq.diff}, the function $M(t)\dfb \max\limits_jx_j(t)$ is non-increasing, so $M(t)\le M(0)$.
\end{lem}

Whereas the class of unbounded solutions of~\eqref{eq.diff} is very broad, under some assumptions on the graph all its \emph{bounded} solutions have simple asymptotic properties,
analogous to the solutions of~\eqref{eq.proto}, namely, each bounded solution converges to a consensus equilibrium point $c\ones_N$. In other words, any solution to the inequality~\eqref{eq.diff} is either convergent or unbounded. In this paper, we disclose conditions, ensuring such a \emph{dichotomic} behavior.
\begin{defn}
The differential inequality~\eqref{eq.diff} is called \emph{dichotomic}, if any of its bounded solutions $x(t)$ has a limit $x^0\dfb\lim_{t\to\infty}x(t)$; it is called
\emph{consensus dichotomic}, if all limits $x^0$ are consensus equilibria $x^0=c^0\ones_N$, $c^0\in\r$.
\end{defn}
\begin{rem}\label{rem.converge}
If $x(t)\xrightarrow[t\to\infty]{}c\ones_N$ and $x(t)$ is an absolutely continuous function, then for any $\tau>0$  the function $\mathfrak{D}(t)=-L(t)x(t)-\dot x(t)$ satisfies the following condition
\be\label{eq.convergence}
\int\limits_t^{t+\tau}\mathfrak{D}(s)ds=x(t)-x(t+\tau)-\int\limits_0^{\tau}L(t+s)x(t+s)ds\xrightarrow[t\to\infty]{} 0.
\ee
\end{rem}
The latter statement follows from Assumption~\ref{ass.l1-loc} and the Dominated Convergence Theorem since $L(t+s)x(t+s)\xrightarrow[t\to\infty]{} 0\,\forall s\in [-\tau,0]$ and $\sup_{t\ge 0}|L(t)x(t)|<\infty$.

If the inequality~\eqref{eq.diff} is consensus dichotomic, then the protocol~\eqref{eq.proto} establishes consensus $x(t)\xrightarrow[t\to\infty]{}c\ones_N$, $c=c(x(0))$. The converse is however not valid: consensus in the system~\eqref{eq.proto} \emph{does not} imply even dichotomy of the inequality~\eqref{eq.diff}, as demonstrated by the following example.
\begin{exmp}\label{ex.trivial}
Consider the differential inequalities
\be\label{eq.exam1}
\begin{aligned}
&\dot x_1\le x_2-x_1,\quad \dot x_2\le 0.
\end{aligned}
\ee
Obviously, a pair of functions $x_1(t)=\sin t$ and $x_2(t)\equiv C\ge 2$ is a bounded yet non-convergent solution to the system~\eqref{eq.exam1}, whereas the corresponding Laplacian flow~\eqref{eq.proto} converges to consensus $x_1(t)\xrightarrow[t\to\infty]{} 0,x_2(t)\equiv 0$.
\end{exmp}

It is well known that the protocol~\eqref{eq.proto} with a static graph $G(t)\equiv G$ establishes consensus if and only if $G$ is \emph{quasi-strongly} connected, or,
equivalently, its Laplacian $L(t)\equiv L$ has a simple eigenvalue at $0$~\citep{AgaevChe:2005,RenBeardBook}.
The necessary and sufficient \emph{dichotomy} condition for the inequality~\eqref{eq.diff} is as follows.
\begin{thm}\label{thm.1}
For a static graph $G(t)\equiv G$, the inequality~\eqref{eq.diff} is \emph{consensus dichotomic} if and only if $G$ is strongly connected.
Otherwise,~\eqref{eq.diff} is \emph{dichotomic} if and only if the strongly connected components of $G$ are isolated, that is, no pair of nodes from different components are connected.
\end{thm}

In the time-varying graph case, the strong connectivity condition has to be replaced by its ``uniform'' version. The \emph{union} of the graphs $G(t)$ over a set $\Delta\subset [0;\infty)$ is
\[
\bigcup\nolimits_{t\in\Delta}G(t)\dfb G\left[\int\nolimits_{\Delta}A(s)\,ds\right].
\]
\begin{defn}\label{def.usc}
A time-varying graph $G(t)$ is uniformly strongly connected (USC) if there exist two numbers $T>0$ and $\delta>0$, such that each union of the graphs $\bigcup_{s\in [t;t+T]}G(s)$ (where $t\ge 0$) is strongly $\delta$-connected.
\end{defn}

The ``limit case'' of the USC condition as $T,\delta\to\infty$ is referred to as the \emph{infinite} strong connectivity (ISC).
\begin{defn}\label{def.isc}
A time-varying graph $G(t)$ is infinitely strongly connected (ISC) if the infinite union of the graphs $\bigcup_{s\ge 0}G(s)$ is strongly $\infty$-connected. More formally, the graph $G=(V_N,E_{\infty})$ is strongly connected, where $E_{\infty}\dfb\{(i,j):\int_0^{\infty}a_{ji}(t)dt=\infty\}$.
\end{defn}

The next theorem extends the consensus dichotomy criterion from Theorem~\ref{thm.1} to the case of time-varying graph.
\begin{thm}\label{thm.2}
For consensus dichotomy of the inequality~\eqref{eq.diff} the graph's $G(t)$ \emph{uniform} strong connectivity is sufficient and its \emph{infinite} strong connectivity
is necessary.
\end{thm}
In the case of static graph, necessary and sufficient conditions boil down to the strong connectivity of the graph. In general, a gap between necessary and sufficient conditions for the consensus dichotomy remains. A similar gap exists between necessary and sufficient conditions for consensus in the network~\eqref{eq.proto}. The assumption of uniform quasi-strong connectivity\footnote{The definitions of uniform and infinite quasi-strong connectivity (UQSC/IQSC) may be obtained from Definitions~\ref{def.usc} and~\ref{def.isc}, replacing the word ``strongly'' by ``quasi-strongly''.},
usually adopted to provide consensus~\citep{Moro:04}, is in fact not necessary, unless one requires additionally the uniform or exponential convergence~\citep{LinFrancis:07,ShiJohansson:13}; the most general necessary condition for consensus is infinite quasi-strong connectivity~\citep{MatvPro:2013}.

At the same time, for some special case of \emph{cut-balanced} interaction graphs the integral connectivity becomes not only necessary but in fact also \emph{sufficient} condition for the consensus dichotomy. We start with a definition.
\begin{defn}\label{def.cut-balance}
The graph $G(t)$ is called \emph{cut-balanced} if a constant $K\ge 1$ exists such that for any subset of nodes $S\subset\{1,\ldots,N\}$ and any $t\ge 0$
the inequalities hold
\ben
K^{-1}\sum_{j\in S}\sum_{k\not\in S}a_{kj}(t)\le
\sum_{j\in S}\sum_{k\not\in S}a_{jk}(t)\le
K\sum_{j\in S}\sum_{k\not\in S}a_{kj}(t). \een
\end{defn}

The class of cut-balanced graphs includes weight-balanced graphs ($\sum_ja_{ij}=\sum_ja_{ji}\,\forall i$), undirected graphs ($a_{ij}(t)=a_{ji}(t)$) and bidirectional or ``type-symmetric''~\citep{TsiTsi:13,MatvPro:2013} graphs, whose weights satisfy
the condition $K^{-1}a_{ji}(t)\le a_{ij}(t)\le Ka_{ji}(t)$ for any $i,j$ and $t\ge 0$. Some other examples can be found in~\citep{TsiTsi:13,ShiJohansson:13-1}.
Under the assumption of cut balance, consensus dichotomy in~\eqref{eq.diff} appears to be equivalent to consensus in the network~\eqref{eq.proto} \citep{TsiTsi:13,MatvPro:2013}.
\begin{thm}\label{thm.3}
Let Assumption~\ref{ass.l1-loc} hold and $G(t)$ be cut-balanced. Then the inequality~\eqref{eq.diff} is dichotomic;
furthermore, the functions $a_{ij}(x_j-x_i)$, $\dot x_i$ and $\mathfrak{D}(t)=-L(t)x(t)-\dot x(t)$ belong to $L_1[0,\infty]$.
The inequality~\eqref{eq.diff} is consensus dichotomic if and only if the graph is ISC.
\end{thm}

Note that the criteria of dichotomy and consensus dichotomy from Theorems~\ref{thm.1},~\ref{thm.2} and~\ref{thm.3} do not allow to estimate the convergence rate for a solution of~\eqref{eq.diff}. This problem is open and seems to be quite non-trivial. However, for some special solutions the convergence rate can be found. In the examples we use one result of this type.
\begin{thm}\label{thm.4}
 Let the graph $G(t)$ be uniformly \emph{quasi-strongly} connected and have a ``leader'' node $s$, such that $a_{sj}(t)\equiv 0\,\forall j$. Then any solution of~\eqref{eq.diff} such that $x_i(t)\ge x_s(0)\,\forall i\,\forall t\ge 0$ \emph{exponentially} converges to $x_s(0)\ones_N$.
\end{thm}

The ``leader'' agent affects the remaining agents, being independent of them. Since $\dot x_s(t)\le 0$ due to~\eqref{eq.diff} and $x_s(t)\ge x_s(0)$, in fact
one has $x_s(t)\equiv x_s(0)$, so the leader's state is invariant. The exponential convergence rate can be found explicitly, as can be seen from the proof.
Notice that the uniform quasi-strong connectivity does not imply neither uniform, nor even integral \emph{strong} connectivity, so the assumptions of Theorem~\ref{thm.4} \emph{do not} imply the consensus dichotomy of~\eqref{eq.diff}: only \emph{some} bounded solutions converge to consensus.

Finally, it should be noticed that the theory, developed in this section, is applicable to the inequalities
\be\label{eq.diff1}
\dot x(t)\ge -L(t)x(t),\,t\ge 0
\ee
without significant changes: if $x(t)$ is a solution to the inequality~\eqref{eq.diff1}, then $(-x(t))$ obeys the inequality~\eqref{eq.diff}, and vice versa.
Lemma~\ref{lem.above} implies that any solution of~\eqref{eq.diff1} is bounded from below. The definitions of dichotomy and consensus dichotomy in~\eqref{eq.diff1} are
the same as for~\eqref{eq.diff}.

\begin{rem}\label{rem.nonlin}
Many results on consensus in linear networks~\eqref{eq.proto} can be extended, without significant changes, to nonlinear consensus algorithms~\citep{Muenz:11,LinFrancis:07,MatvPro:2013,Moro:05}, which in turn may be associated with nonlinear counterparts of the inequality~\eqref{eq.diff}. Theorems~\ref{thm.1},\ref{thm.2},\ref{thm.3} and \ref{thm.4} can be extended to the nonlinear case, however, we confine ourselves to the linear inequalities~\eqref{eq.diff} due to the page limit.
\end{rem}

\section{Examples and Applications}\label{sec.example}

In this section, the results from Section~\ref{sec.ineq1} are used to derive some recent results on multi-agent coordination in a unified way, and also extend them by discarding some technical assumptions (e.g. the dwell-time positivity).

\subsection{Target aggregation and containment control}

Consider a team of $N$ mobile agents, obeying the single integrator model $\dot\xi_i(t)=u_i(t)\in\r^d$, $i\in V_N$, where
$\xi_i(t)$ stands for the position of agent $i$ and $u_i(t)$ is its velocity, being also the \emph{control} input.
The agents' cooperative goal, sometimes called the \emph{target aggregation}~\citep{ShiHong:09}, is to gather within some fixed \emph{target set} $\Omega\subseteq\r^d$, which is assumed to be \emph{convex} and \emph{closed}.

Were the set $\Omega$ known by all of the agents, to gather in it would be a trivial problem. However, the knowledge about $\Omega$, in general, is available only to a few \emph{informed} agents (whose set may evolve over time), whereas the remaining agents can obtain the information about the desired set only via communication over some graph (generally, time-varying). We examine a distributed protocol, similar to that proposed in~\citep{ShiHong:09}
\be\label{eq.aggreg}
\dot \xi_i(t)=\sum_{j=1}^Na_{ij}(t)(\xi_j(t)-\xi_i(t))+a_{i0}(t)[\omega_i(t)-\xi_i(t)].
\ee
Here $i\in V_N$, the matrix $A(t)=a_{ij}(t)$ describes the (weighted) interaction graph and the gains $a_{i0}(t)\ge 0$ are responsible for the attraction to the target set $\Omega$. Agent $i$ is \emph{informed}\footnote{Our terminology differs from~\citep{ShiHong:09}, where the set of informed agents if static, but the target  is accessible to them only at some time instants. We call the agent informed at time $t\ge 0$ if it is aware of some element $\omega_i(t)\in\Omega$.} at time $t\ge 0$ if $a_{i0}(t)>0$, in this case $\omega_i(t)\in\Omega$; otherwise the choice of $\omega_i(t)$ can be arbitrary.

Let $P_{\Omega}$ be the operator of projection onto $\Omega$; as in Section~\ref{sec.prelim}, we denote $\xi^p\dfb\xi-P_{\Omega}(\xi)$. Let $x_i(t)\dfb\frac{1}{2}|\xi_i^p(t)|^2$. Notice that $\xi_j^{p}(t)^{\top}\xi_i^{p}(t)\le x_i(t)+x_j(t)$ and hence $\xi_j^{p}(t)^{\top}\xi_i^{p}(t)-|\xi_i^p(t)|^2\le x_j(t)-x_i(t)$. Using~\eqref{eq.gradient}, one has
\be\label{eq.ineq-ex1}
\begin{aligned}
\dot x_i(t)&\overset{\eqref{eq.gradient}}{=}\dot\xi_i(t)^{\top}\xi_i^p(t)\overset{\eqref{eq.aggreg}}{=}\sum_{j=1}^Na_{ij}(t)(\xi_j(t)-\xi_i(t))^{\top}\xi_i^p(t)-\\
&\phantom{\dot\xi_i(t)^{\top}\xi_i^p(t)=}-a_{i0}(t)(\omega_i(t)-\xi_i(t))^{\top}\xi_i^p(t)
\overset{\eqref{eq.projection1},\eqref{eq.projection}}{\le}\\
&\le \sum_{j=1}^Na_{ij}(t)(\xi_j^p(t)-\xi_i^p(t))^{\top}\xi_i^p(t)-a_{i0}(t)|\xi_i^p(t)|^2\le \\
&\le \sum_{j=1}^Na_{ij}(t)(x_j(t)-x_i(t))-2a_{i0}(t)x_i(t)
\end{aligned}
\ee
for almost all $t\ge 0$. Since $a_{i0}x_i(t)\ge 0$, $x(t)=(x_1(t),\ldots,x_N(t))^{\top}\ge 0$ is a solution to~\eqref{eq.diff}.

In order to formulate the convergence criterion, it is convenient to consider the target set $\Omega$ as a ``virtual agent''~\citep{ShiHong:09}, indexed by $0$, and introduce the extended matrix
$\hat A(t)=(a_{ij}(t))_{i,j=0}^N$, where $a_{0j}\equiv 0\,\forall j$ and other $a_{ij}$ are the weights from~\eqref{eq.aggreg}. 
\begin{thm}\label{thm.aggreg}
Suppose that Assumption~\ref{ass.l1-loc} holds for the extended matrix $\hat A(t)$ and one of the conditions is valid
\begin{enumerate}
\item the ``extended'' graph $\hat G(t)=G[\hat A(t)]$ is uniformly quasi-strongly connected;
\item the graph $G(t)=G[A(t)]$ is cut-balanced and ISC, and also $\sum_{i=1}^N\int_0^{\infty}a_{i0}(t)dt=\infty$.
\end{enumerate}
Then the agents converge to $\Omega$ in the sense that $x_i(t)\xrightarrow[t\to\infty]{}0$; in case (1) the convergence is exponential.
\end{thm}
\pf
The second part of Theorem~\ref{thm.aggreg} is immediate from~\eqref{eq.ineq-ex1} and Theorem~\ref{thm.3}\footnote{Note that Theorem~\ref{thm.3} is applied to the \emph{original} graph $G(t)$, while the extended graph $\hat G(t)$ is not cut-balanced.}. As has been mentioned, $a_{i0}x_i(t)\ge 0$, and therefore~\eqref{eq.ineq-ex1} implies the differential inequality~\eqref{eq.diff}. Thus the functions $x_i(t)$ converge to a consensus value
$x_i(t)\xrightarrow[t\to\infty]{}x_*$ and $\dot x_i$, $a_{ij}(x_j-x_i)$ and $\mathfrak{D}(t)=-\dot x(t)-L(t)x(t)$ are $L_1$-summable. Since ${\mathfrak D}_i(t)\ge 2a_{i0}(t)x_i(t)\ge 0$, the functions $a_{i0}x_i$  also $L^1$-summable.
If $x_*>0$, then $a_{i0}$ is $L^1$-summable for any $i$, which contradicts to the assumption $\sum_{i=1}^N\int_0^{\infty}a_{i0}(t)dt=\infty$. Hence $x_*=0$, which proves the second statement.

Introducing the additional function $x_0(t)\equiv 0$, \eqref{eq.ineq-ex1} implies that the extended vector $\hat x=(x_0,\ldots,x_N)^{\top}$ satisfies the inequality
$\dot{\hat x}(t)\le -L[\hat A(t)]\hat x(t)$. The first part of Theorem~\ref{thm.aggreg} now follows from Theorem~\ref{thm.4} (applied to $\hat A$ and $s=0$), recalling that $x_i(t)\ge x_0(t)\equiv 0$ for any $i$.\epf

Theorem~\ref{thm.aggreg} extends the results from Theorems~15 and~17 from~\citep{ShiHong:09}. Unlike~\citep{ShiHong:09},  the matrix $\hat A(t)$ need not be piecewise-constant with positive dwell time between its consecutive switchings, and  the weights $a_{ij}(t)$ do not need to be uniformly strictly positive. In case~(1) our result also ensures exponential convergence, which is not directly implied by the results of~\citep{ShiHong:09}. At the same time, the paper~\citep{ShiHong:09} deals with a more general protocol, where the terms $(\omega_i(t)-\xi_i(t))$ in~\eqref{eq.aggreg} are replaced by the nonlinearities $f_i(\xi_i,t)$, satisfying the condition
$$
(\xi^p)^{\top}f_i(\xi,t)\le -\varkappa_i(|\xi^p|)\quad\forall\xi\in\r^d,
$$
where $\varkappa_i(\cdot)$ is a $\mathcal{K}$-function. The second part of Theorem~\ref{thm.aggreg} (corresponding to Theorem~17 in~\citep{ShiHong:09}) retains its validity for this general case, as can be seen from its proof. The extension of the first part of Theorem~\ref{thm.aggreg} (but for the exponential stability, which in general fails) to the algorithm from~\citep{ShiHong:09} is beyond the scope of this paper, since it requires a nonlinear version of Theorem~\ref{thm.4} (see Remark~\ref{rem.nonlin}).

A special case of the target aggregation problem is the \emph{containment} control problem with static leaders~\citep{RenCaoBook}, where the desired set $\Omega=\mathrm{conv}\{\xi_{N+1},\ldots,\xi_{N+q}\}$ is a convex polytope, spanned by the fixed vectors $\xi_{N+1},\ldots,\xi_{N+q}$. These vectors are considered as the positions of $q\ge 1$ static agents, called \emph{leaders}. Only a few ``informed'' agents are aware of the position of one or several leaders. In order to gather the agents in the set $\Omega$, the consensus-like protocol has been proposed~\citep{RenCaoBook}
\be\label{eq.contain}
\dot \xi_i(t)=\sum_{j=1}^{N+q}a_{ij}(t)(\xi_j(t)-\xi_i(t)).
\ee
Introducing the gains $a_{i0}(t)$ and vectors $\omega_i(t)$ as follows
\[
a_{i0}(t)=\sum_{j=N+1}^{N+q}a_{ij}(t),\quad \omega_i(t)    =
\begin{cases}
\sum\limits_{j=N+1}^{N+q}\frac{a_{ij}(t)}{a_{i0}(t)}\xi_j\in\Omega\\
0,\quad\text{otherwise},
\end{cases}
\]
the protocol~\eqref{eq.contain} becomes a special case of the more general aggregation algorithm~\eqref{eq.aggreg}.
The first part of Theorem~\ref{thm.aggreg} extends the result of Theorem~5.3 in \citep{RenCaoBook}, relaxing the connectivity assumptions.

In general, target aggregation and containment control algorithms do not lead to the consensus of agents; however, asymptotic consensus can be established if the informed agents are able to compute the projection of their position onto $\Omega$ and one may take $\omega_i(t)=P_{\Omega}(\xi_i(t))$.
\begin{lem}\label{lem.conse}
Under the assumptions of Theorem~\ref{thm.aggreg}, the protocol~\eqref{eq.aggreg} with $\omega_i(t,\xi_i)=P_{\Omega}(\xi_i(t))$ provides
\be\label{eq.conse}
\lim_{t\to\infty}|\xi_i(t)-\xi_j(t)|=0\quad\forall i,j.
\ee
\end{lem}
\pf
The condition (1) in Theorem~\ref{thm.aggreg} entails that $\xi_i^p(t)=\xi_i-P_{\Omega}(\xi_i(t))\to 0$ and hence $f_i(t)=a_{i0}(t)(\omega_i(t)-\xi_i(t))\to 0$ as $t\to\infty$. Rewriting~\eqref{eq.aggreg} as the ``disturbed'' consensus dynamics~\eqref{eq.proto}
\be\label{eq.proto-disturb}
\dot\xi_i(t)=\sum_{j=1}^Na_{ij}(t)(\xi_j(t)-\xi_i(t))+f_i(t)\quad\forall i,
\ee
the statement follows from the robust consensus criterion~\cite[Proposition~4.8]{ShiJohansson:13}. If the condition (2) in Theorem~\ref{thm.aggreg} holds, then the functions $f_i(t)$ are $L_1$-summable, and consensus~\eqref{eq.conse} follows from~\cite[Proposition~5.3]{ShiJohansson:13}.
\epf

Containment control and target aggregation control with \emph{time-varying}  target sets~\citep{RenCaoBook} are
beyond the scope of this paper; these problems require the extensions of Theorems~\ref{thm.3} and~\ref{thm.4} to ``disturbed'' differential inequalities, associated to protocols~\eqref{eq.proto-disturb}.

\subsection{Optimal consensus and distributed optimization}

The result of Lemma~\ref{lem.conse} has been extended in~\citep{ShiJohanssonHong:13} to the case where agents
cannot find any element of the target set $\Omega$, representable as an intersection of several convex closed sets $\Omega=\bigcap_{i=1}^N\Omega_i\ne\emptyset$. Agent $i$ is aware of the set $\Omega_i$ and is able to calculate the projection of its state onto this set; the other sets $\{\Omega_j\}_{j\ne i}$ are unavailable to it. The common goal of the agents is to reach consensus at some point from $\Omega$. In~\citep{ShiJohanssonHong:13} this goal was called \emph{optimal consensus} since the problem of convex set intersection is dual to the distributed convex optimization problem~\cite[Section~1]{ShiJohanssonHong:13}.

In order to establish this optimal consensus, the following protocol has been proposed in~\citep{ShiJohanssonHong:13}
\be\label{eq.optim}
\dot \xi_i(t)=\sum_{j=1}^Na_{ij}(t)(\xi_j(t)-\xi_i(t))+P_{\Omega_i}(\xi_i(t))-\xi_i(t).
\ee
Although the protocol~\eqref{eq.optim} is similar to~\eqref{eq.aggreg}, the conditions for its convergence appear to be more restrictive.
The following theorem extends Lemma~4.3 and Theorems~3.1 and~3.2 in~\citep{ShiJohanssonHong:13}.
\begin{thm}\label{thm.optimal}
Let $\Omega_i$ be closed convex sets and $\Omega=\cap_{i=1}^N\Omega_i\ne\emptyset$. Suppose that Assumption~\ref{ass.l1-loc} holds and one of the following two conditions is valid
\begin{enumerate}
\item the graph $G[A(\cdot)]$ is USC;
\item the graph $G[A(\cdot)]$ is cut-balanced and ISC.
\end{enumerate}
Then the protocol~\eqref{eq.optim} provides \emph{equidistant} deployment of the agents with respect to the set $\Omega$, that is,
the limit $\lim\limits_{t\to\infty}|\xi_i(t)-P_{\Omega}(\xi_i(t))|=d_*\ge 0$ exists and is independent of $i$. If $\Omega$ is bounded, then the agents converge to $\Omega$ (that is, $d_*=0$) and reach consensus~\eqref{eq.conse}.
\end{thm}
\pf
We denote $\xi_i^p(t)=\xi_i(t)-P_{\Omega}(\xi_i(t))$, $x_i(t)\dfb\frac{1}{2}|\xi_i^p(t)|^2$ and $y_i(t)\dfb d_{\Omega_i}(\xi_i(t))=|P_{\Omega_i}(\xi_i(t))-\xi_i(t)|^2$. Applying~\eqref{eq.projection1} to $\Omega=\Omega_i$, $\omega=P_{\Omega}(\xi_i)\in\Omega\subseteq\Omega_i$ and $\xi=\xi_i$, one shows that $(\xi_i^p)^{\top}(P_{\Omega_i}(\xi_i)-\xi_i)\le -y_i\le 0$.
Similar to the proof of~\eqref{eq.ineq-ex1}, it can be shown that
\be\label{eq.ineq-ex2}
\dot x_i(t)\le \sum_{j=1}^Na_{ij}(t)(x_j(t)-x_i(t))-y_i(t)\quad\forall i.
\ee
Theorems~\ref{thm.2} and~\ref{thm.3} imply the first statement: $x_i(t)$ converge
to a common value $x_*$ as $t\to\infty$. To prove the second statement, it suffices to show that $y_i(t)\to 0$ as $t\to\infty$ for any $i$; in this case, consensus~\eqref{eq.conse} follows from the results of~\citep{ShiJohansson:13} (see the proof of Lemma~\ref{lem.conse}). Since $0\le y_i(t)\le |\xi_i^p(t)|^2\xrightarrow[t\to\infty]{} 2x_*$, consensus is established if $x_*=0$. Otherwise, one may notice that $x_i\ge 0$ are bounded functions due to Lemma~\ref{lem.above} and thus $\xi_i,\dot \xi_i,\dot y_i$ are also bounded. Using Remark~\ref{rem.converge}, the convergence $y_i(t)\xrightarrow[t\to\infty]{} 0$ is proved by standard Barbalat-type arguments, since $0\le y_i(t)\le\mathfrak{D}_i(t)$ due to~\eqref{eq.ineq-ex2}.
\epf

Comparing Theorem~\ref{thm.optimal} with the results of~\citep{ShiJohanssonHong:13}, one notices that two restrictions are discarded: the positive dwell-time between switchings of the matrix $A(t)$ and the uniform positivity of its non-zero entries.

\subsection{Target set surrounding and the Altafini model.}

Another extension of the target aggregation problem has been addressed in~\citep{LouHong:15}. This problem deals with the planar ($d=2$) motion of $N$ mobile agents, whose positions are represented by complex numbers $\xi_i\in\C$. A closed convex set $\Omega\subseteq\C$ is given. The agents' motion is described by the following equations
\be\label{eq.surround}
\dot \xi_i(t)=\sum_{j=1}^Na_{ij}(t)(w_{ij}(t)\xi_j^p(t)-\xi_i^p(t))\quad\forall i.
\ee
Here, as in the previous subsections, $\xi_i^p\dfb\xi_i-P_{\Omega}(\xi_i)$ and $A(t)=(a_{ij}(t))$ stands for the weighted adjacency matrix. Besides this, the protocol~\eqref{eq.surround}
employs complex-valued matrix $W(t)=(w_{ij}(t))$, whose entries belong to the unit circle $|w_{ij}(t)|=1$. The following lemma is proved similarly to the first statement in Theorem~\ref{thm.optimal}.
\begin{lem}\label{lem.surround}
If Assumption~\ref{ass.l1-loc} holds and one of the conditions (1) or (2) from Theorem~\ref{thm.optimal} is valid, then the protocol~\eqref{eq.surround} renders the agents equidistant from $\Omega$: there exists a limit $d_*=\lim\limits_{t\to\infty}|\xi_i^p(t)|\ge 0$, independent of $i$.
\end{lem}
\pf
Similar to the proofs of Theorems~\ref{thm.aggreg} and~\ref{thm.optimal}, introduce the functions $x_i(t)\dfb\frac{1}{2}|\xi_i^p(t)|^2$ and note that
\be\label{eq.ineq-ex3}
\begin{aligned}
\re \xi_i^p(t)^*&w_{ij}(t)\xi_j^p(t)\le \left|\xi_i^p(t)^*w_{ij}(t)\xi_j^p(t)\right|=\\
&=|\xi_i^p(t)|\,|\xi_j^p(t)|\le x_i(t)+x_j(t).
\end{aligned}
\ee
Using~\eqref{eq.gradient} and retracing arguments from~\eqref{eq.ineq-ex1}, one has
\be\label{eq.ineq-ex3+}
\begin{aligned}
\dot x_i(t)\overset{\eqref{eq.gradient}}{=}\langle\xi_i^p(t),\dot\xi_i(t)\rangle=\re\xi_i^p(t)^*\dot\xi_i(t)\overset{\eqref{eq.surround}}{=}\\ =\sum_ja_{ij}(t)\left(\re\xi_i^p(t)^*w_{ij}(t)\xi_j^p(t)-|\xi_i^p(t)|^2\right)\overset{\eqref{eq.ineq-ex3}}{\le} \\ \le\sum_ja_{ij}(t)(x_j(t)-x_i(t))
\end{aligned}
\ee
(here $\langle z_1,z_2\rangle=\re z_1\re z_2+\im z_1\im z_2=\re z_1^*z_2$ is the inner product in $\C\cong\r^2$).
Therefore, $x(t)=(x_1(t),\ldots,x_N(t))^{\top}\ge 0$ is a solution to~\eqref{eq.diff}. The statement of lemma now follows from Theorems~\ref{thm.2} and~\ref{thm.3}.
\epf

\subsubsection{Target surrounding}
In the remainder of this section, we consider two special cases of the dynamics~\eqref{eq.surround}.
The first special case is the \emph{surrounding} problem from~\citep{LouHong:15},
dealing with the case of static $W(t)\equiv W$. It is said that the agents \emph{surround} the target set $\Omega$ if $w_{ij}\xi_j^p(t)-\xi_i^p(t)\xrightarrow[t\to\infty]{} 0$.
If $d_*>0$, this means that the complex argument of $w_{ij}$ determines the angle between the vectors $\xi_j^p(t)$ and $\xi_i^p(t)$ for large $t\ge 0$.
As discussed in~\citep{LouHong:15}, the target surrounding with $d_*>0$ is usually possible only for \emph{consistent} matrices $W$, which means that $w_{ij}=p_i^*p_j$, where $p_1,\ldots,p_N$ are complex numbers with $|p_i|=1$. Lemma~\ref{lem.surround} enables to extend statement (i) in~\cite[Theorem~2]{LouHong:15} as follows.
\begin{thm}\label{thm.surround}
Suppose that the graph $G[A(\cdot)]$ is USC and Assumption~\ref{ass.l1-loc} holds. Let $W(t)\equiv W$ be consistent, i.e. $w_{ij}=p_i^*p_j$. Then the protocol~\eqref{eq.surround} provides the target set surrounding $p_i\xi_i^p(t)-p_j\xi_j^p(t)\xrightarrow[t\to\infty]{} 0$.
\end{thm}
\pf
Introducing the function $\mathfrak{D}(t)=-\dot x(t)-L(t)x(t)$,~\eqref{eq.ineq-ex3+} entails that
\be
\begin{aligned}
\mathfrak D_i(t)&=\sum_{j=1}^Na_{ij}(t)\left[x_i(t)+x_j(t)-\re\xi_j^p(t)^*w_{ij}\xi_i^p(t)\right]\\
&=\frac{1}{2}\sum_{j=1}^Na_{ij}(t)\left|p_j\xi_j^p(t)-p_i\xi_i^p(t)\right|^2.
\end{aligned}
\ee
Recalling that $\dot \xi_i(t)$ are bounded functions, and hence $\xi_i(t),\xi_i^p(t)$ are Lipschitz and applying~\eqref{eq.convergence} to $\tau=T$, where $T$ is the period from Definition~\ref{def.usc}, it is now easy to prove that $p_j\xi_j^p(t)-p_i\xi_i^p(t)\xrightarrow[t\to\infty]{} 0$ in a way similar to the proof of Theorem~5 in~\citep{Pro13AutDelay}.
\epf
\begin{rem}
Although the explicit computation of $d_*$ is non-trivial, sufficient conditions for its positivity have been offered in~\citep{LouHong:15}.
\end{rem}

Unlike the result from~\citep{LouHong:15}, Theorem~\ref{thm.surround} is applicable to the weighted interaction graph, discarding the restriction of the dwell-time existence.
In the case where $\Omega=\{0\}$ is a singleton, the target surrounding implies that the agents converge to a circular formation~\citep{LouHong:15}, or reach ``complex consensus''~\citep{DongQui:15}. As shown in~\citep{ProCao16-4}, in this special case Theorem~\ref{thm.surround} retains its validity, relaxing the USC condition to the uniform \emph{quasi-strong} connectivity.

\subsubsection{The Altafini model of opinion formation}

The central problem in opinion formation modeling is to elaborate
models of opinion evolution in social networks that are able to explain both consensus of opinions and their persistent disagreement. The recent models, proposed in the literature,
explain this disagreement by presence of ``prejudiced'' or ``informed'' agents~\citep{Friedkin:2015,XiaCao:11}, influenced by some constant external factors,
and homophily effects, such as bounded confidence~\citep{Krause:2002,Blondel:2009} and biased assimilation~\citep{Dandekar:2013}.
Another type of opinion dynamics has been proposed in~\citep{Altafini:2012,Altafini:2013}. This model describes the mechanism of bi-modal polarization, or ``bipartite consensus'' in a \emph{signed}
or \emph{coopetition}~\citep{HuZheng:2014} network with mixed positive and negative ties.

The Altafini model is a special case of the dynamics~\eqref{eq.surround}, where $\Omega=\{0\}$, $\xi_i(t)=\xi_i^p(t)\in\r$ and $w_{ij}(t)\in\{1,-1\}$.
Denoting $b_{ij}=a_{ij}w_{ij}$, the Altafini model is as follows
\be\label{eq.altafini}
\dot\xi_i(t)=\sum_{j=1}^N\left[b_{ij}(t)\xi_j(t)-|b_{ij}(t)|\xi_i(t)\right]\quad\forall i.
\ee
The coupling term $(b_{ij}(t)\xi_j(t)-|b_{ij}(t)|\xi_i(t))$ (infinitesimally) drives $\xi_i$ to $\xi_j$ when
$b_{ij}>0$ and to $-\xi_j$ if $b_{ij}<0$.

Lemma~\ref{lem.surround} implies the following important corollary, combining the results of Theorems~2 and~3 in~\citep{ProMatvCao:2016} and discarding the restrictive
\emph{digon-symmetry} assumption $b_{ij}b_{ji}\ge 0$, adopted in~\citep{Altafini:2013,ProMatvCao:2016}.
\begin{cor}
If the matrix $A(t)=(|b_{ij}(t)|)$ satisfies the assumptions of Lemma~\ref{lem.surround}, the protocol~\eqref{eq.altafini} establishes~\emph{modulus consensus}: the limit
$x_*\dfb\lim\limits_{t\to\infty}|\xi_i(t)|\ge 0$ exists and is independent of $i$.
\end{cor}

As shown in~\citep{ProMatvCao:2016}, modulus consensus implies either asymptotic \emph{stability} (for any initial condition the solution converges to $0$) or \emph{polarization}:
the community is divided into two ``hostile camps'', reaching consensus at the opposite opinions $x_*$ and $-x_*$ (here $x_*$ is non-zero for almost all initial conditions).
In the case where $B(t)\equiv B$ is constant and the graph $G[A]\equiv G$ is strongly connected, polarization (```bipartite consensus'') is equivalent to the \emph{structural balance}
of the network~\citep{Altafini:2013}; the extensions of the latter result to more general static and special switching graphs can be found in~\citep{Meng:16-2,ProMatvCao:2014,ProCao:2014,ProMatvCao:2016,LiuChenBasar:2018}. Whether protocol provides polarization for a general matrix $B(t)$ seems to be a non-trivial open problem. Numerous extensions of the Altafini model~\eqref{eq.altafini} have been proposed recently, see e.g.~\citep{Valcher:14,XiaCaoJohansson:16,LiuChenBasar:2018,MengShiCao:16}.

\section{Proofs of the Main Results}\label{sec.proof}

Henceforth Assumption~\ref{ass.l1-loc} is supposed to be valid. The proofs of the main results employ the useful construction of \emph{ordering}, used in analysis of usual consensus algorithms~\citep{TsiTsi:13,MatvPro:2013,ProMatvCao:2016}. Given $N$ functions $x_1(t),\ldots,x_N(t)$, let  $[k^1(t), \ldots, k^N(t)]$ be {\it the ordering permutation}, sorting the set $\{x_1(t),\ldots, x_N(t)\}$ in the ascending order.
Precisely, the inequalities hold
\be\label{eq.ordering}
y_1(t)\le y_2(t)\le\ldots\le y_N(t),\quad y_i\dfb x_{k^i(t)}(t).
\ee
Here the time $t$ may be continuous ($t\in [0;\infty)$) or discrete ($t=0,1,\ldots$). Obviously, the indices $k^i(t)$ may be defined non-uniquely.
If $x_i(t)$ are absolutely continuous on $[0;\infty)$, then one always can choose $k^i(t)$ to be measurable; moreover, $y_i(t)$ are absolutely continuous and
\be\label{eq.ordering-diff}
\dot y_i(t)=\dot x_{k^i(t)}(t)\quad\forall i.
\ee
for almost all $t\ge 0$. This is implied by Proposition~2 in~\citep{TsiTsi:13Extended} where the constructive procedure of choosing $k^i(t)$ is described.

We will use the following well-known fact.
\begin{lem}~\citep[Sect.~1.2.2]{RenBeardBook}
The Cauchy transition matrix $\Phi(t,t_0)$ of the system~\eqref{eq.proto} is stochastic for any $t\ge t_0$ (its entries thus belong to $[0;1]$).
\end{lem}
The Cauchy formula, applied to the equation
\be\label{eq.aux1}
\dot x(t)=-L(t)x(t)+f(t),\,t\ge t_0,
\ee
yields in $x(t)=\Phi(t,t_0)x(t_0)+\int_{t_0}^{t}\Phi(t,s)f(s)ds$, leading to the following corollaries.
\begin{cor}\label{lem.aux2}
For any solution of~\eqref{eq.aux1}, one has $m(t)\dfb\min_i x_i(t)\geq m(t_0)-\int_{t_0}^{t}\sum_i|f_i(s)|ds$.
\end{cor}
\begin{cor}\label{cor.neg}
If $f(t)\le 0$, then for any solution of~\eqref{eq.aux1} one has $x(t)\le\Phi(t,t_0)x(t_0)$.
\end{cor}

\subsection{Proof of Lemma~\ref{lem.above}}

The proof is immediate from~\eqref{eq.ordering-diff}. Since $M(t)=y_N(t)$, one has $\dot M(t)=\dot x_{k^N(t)}(t)\le \sum_{j=1}^Na_{k^N(t)}(x_j(t)-M(t))\le 0$ and thus $M(t)$ is a non-increasing function.

\subsection{Proof of Theorem~\ref{thm.2}, sufficiency part}

The proof follows the line of the proof of Theorem~2 in~\citep{ProMatvCao:2016}. We first prove the following extension of Lemma~\ref{lem.above}.
\begin{lem}\label{lem.aux}
For any $T\ge 0,\delta>0$ a number $\theta=\theta(\delta,T,A(\cdot))\in (0;1)$ exists such that the following two statements are valid for any solution of~\eqref{eq.diff}
\begin{enumerate}
\item if $\max\limits_i x_i(t_0)=M$ and $x_j(t_0)\le M-\rho$ for some $j$, $t_0\ge 0$ and $\rho\ge 0$, then
$x_j(t_0+T)\le M-\theta\rho$;
\item if, additionally, $\int_{t_0}^{t_0+T}a_{kj}(t)dt\ge\delta$ for some $k$ then $x_k(t_0+T)\le M-\theta\rho$.
\end{enumerate}
\end{lem}
\pf
Let $\Delta\dfb [t_0;t_0+T]$, $s_i(t)\dfb \sum_{m=1}^Na_{im}(t)$ and $S_i(t)=\int_{t_0}^t s_i(\tau)\,d\tau$. Due to the boundedness of $A(t)$,
$S_i(t)\le C\,\forall i\forall t\in\Delta$, where $C=C(T)$ is independent of $t_0$. Since $x_i(t)\le M\,\forall i\,\forall t\in\Delta$ due to Lemma~\ref{lem.above}, we have
\ben
\frac{d}{dt}(M-x_j(t))=-\dot x_j(t)\stackrel{\eqref{eq.diff}}{\ge} -s_j(t)(M-x_j(t))\quad\forall t\in\Delta.
\een
Denoting $\theta_1\dfb e^{-C}$, the latter inequality implies that
\ben
M-x_j(t)\ge e^{-S_j(t)}(M-x_j(t_0))\ge \theta_1\rho\quad\forall t\in\Delta.
\een
By noticing that $x_j(t)-x_k(t)=(M-x_k(t))-(M-x_j(t))$ and denoting $\theta_2\dfb \delta\theta_1^2$ one obtains that
\ben
\frac{d}{dt}(M-x_k(t))\stackrel{\eqref{eq.diff}}{\ge} -s_k(t)(M-x_k(t))+a_{kj}(t)\theta_1\rho\quad\forall t\in\Delta
\een
and therefore $M-x_k(t)\ge \theta_1\rho\int_{t_0}^t e^{-S_k(t-\tau)}a_{kj}(\tau)d\tau\ge\theta_2\rho$. Thus statements~1 and~2 hold for $\theta\dfb \min(\theta_1,\theta_2)$.\qed

\begin{cor}
    Let the graph $G(t)$ be USC with the period $T>0$ and the threshold $\delta>0$ and $\theta=\theta(\delta,T,A(\cdot))$ be the constant from Lemma~\ref{lem.aux}.
    Then the ordering~\eqref{eq.ordering} of any solution to~\eqref{eq.diff} satisfies the inequalities
\be\label{eq.ordering-usc}
y_{m+1}(t+T)\le \theta y_m(t)+(1-\theta)y_N(t)
\ee
where $m=1,\ldots,N-1$ and $t\ge 0$.
\end{cor}
\pf
Introducing the set of indices $S_m(t_0)=\{k^1(t),\ldots,k^m(t)\}$, one has $x_i(t_0)\le y_m(t_0)\,\forall i\in S_m(t_0)$.
Applying Lemma~\ref{lem.aux} to $t_0=t$, $M=y_N(t)$ and $\rho=M-y_m(t_0)$, one shows that
\be\label{eq.aux0}
x_i(t+T)\le \theta y_m(t)+(1-\theta)y_N(t)
\ee
for any $i\in S_m(t)$. By Definition~\ref{def.usc}, there exist nodes $j\in S_m(t_0)$ and $k\not\in S_m(t_0)$ such that $\int_{t_0}^{t_0+T} a_{kj}(s)ds\ge\delta$. Lemma~\ref{lem.aux}  implies that~\eqref{eq.aux0} holds also for $i=k$ and thus $m+1$ different indices $i\in S_m(t_0)\cup\{k\}$ satisfy~\eqref{eq.aux0}. This entails~\eqref{eq.ordering-usc} by definition of the ordering~\eqref{eq.ordering}.
\qed

We are now ready to prove the sufficiency part of Theorem~\ref{thm.2}. Let the graph $G(t)$ be USC. Given a bounded solution to~\eqref{eq.diff}, consider its ordering~\eqref{eq.ordering}. Lemma~\ref{lem.above} implies that $y_N(t)=M(t)$ is non-increasing and thus has a finite limit $M_*\dfb \lim\limits_{t\to\infty}y_N(t)>-\infty$. Using~\eqref{eq.ordering-usc} for $m=N-1$, one shows that $\varliminf\limits_{t\to\infty}y_{N-1}(t)\ge M_*$ and hence $y_{N-1}(t)\xrightarrow[t\to\infty]{} M_*$.
The inequality~\eqref{eq.ordering-usc} for $m=N-2$ implies now that $y_{N-2}(t)\xrightarrow[t\to\infty]{} M_*$, and so on, $y_{1}(t)\xrightarrow[t\to\infty]{} M_*$. Therefore, $x(t)\to M_*\ones_N$.\qed

\subsection{Proof of Theorem~\ref{thm.2}, necessity part}

We are going to show that the consensus dichotomy of~\eqref{eq.diff} implies (under Assumption~\ref{ass.l1-loc}) that the graph $G[A(\cdot)]$ is ISC, that is,
the graph $G_{\infty}=(V_N,E_{\infty})$, introduced in Definition~\ref{def.isc}, is strongly connected. Suppose, on the contrary, that it has multiple strongly connected components. As follows from the results of \cite[Chapter~3]{HararyBook:1965}, at least one of this components is ``closed'' and has no incoming arcs; let $S\subset V_N$ denote the set of its nodes. By assumption, nodes from $S^c$ are not connected to the nodes from $S$ in $G_{\infty}$, and hence $\sum_{i\in S,j\not\in S}\int_{t_0}^{\infty}a_{ij}(\tau)d\tau<1/2$ for sufficiently large $t_0>0$. We are going construct a bounded solution $x(t)$ to~\eqref{eq.diff}, which does not converge to a consensus point.
Define the matrix $\bar A(t)=(\bar a_{ij}(t))$ as follows
\[
\bar a_{ij}(t)=
\begin{cases}
0,\quad &t>t_0\,\&\, i\in S^c\,\&\, j\in S\\
a_{ij}(t),\,&\text{otherwise},
\end{cases}
\]
and let $x(t)$, $t\ge 0$ be the solution to the Cauchy problem
\[
\dot{x}(t)=-L[\bar A(t)]\bar x(t),\quad x_i(t_0)=
\begin{cases}
1,i\in S\\
0,i\in S^c.
\end{cases}
\]
Obviously, $x(t)$ is a solution to~\eqref{eq.proto} for $t\in [0;t_0]$. When $t>t_0$, one has
$\bar x_i(t)\equiv 0$ for $i\in S^c$ and $\bar x_i(t)\in [0;1]$ when $i\in S$. Hence obeys~\eqref{eq.diff} when $t>t_0$. Indeed, $\dot x_i(t)=0\le \sum_{j}a_{ij}(t)(x_j(t)-x_i(t))$ when $i\in S^c$ and $\dot x_i(t)=\sum_{j}a_{ij}(t)(x_j(t)-x_i(t))$ for $i\in S$. To prove that $x(t)$ does not converge to a consensus equilibrium,
we are now going to show that $x_i(t)\ge 1/2$ when $i\in S$ and $t>t_0$.

Consider the matrix $\tilde A(t)=(a_{ij}(t))_{i,j\in S}$ and let $\tilde L(t)=L[\tilde A(t)]$ be the corresponding Laplacian matrix. The truncated vector $\tilde x(t)=(\bar x_i(t))_{i\in S}$ satisfies the equation
\[
\dot{\tilde x}(t)=-\tilde L(t)\tilde x(t)-\tilde f(t),
\]
where $\tilde f_i(t)=\sum_{j\in S^c}a_{ij}(t)(x_j(t)-x_i(t))$ for any $i\in S$, and hence $\int_{t_0}^{\infty}\sum_i|f_i(t)|dt<1/2$.
Applying Corollary~\ref{lem.aux2}, one shows that $\min_{i\in S} x_i(t)\ge 1/2\,\forall t>t_0$. The contradiction proves that $G_{\infty}$ is strongly connected.\qed

\subsection{Proof of Theorem~\ref{thm.1}}

The statements about \emph{consensus dichotomy} follow from the more general Theorem~\ref{thm.2}.
Assume that the graph $G$ has $s>1$ strongly connected components $G_1,\ldots,G_s$. We are going to prove that
the inequality~\eqref{eq.diff} is dichotomic if and only if these components are isolated.
The sufficiency is immediate from the consensus dichotomy criterion: if $G_i$ are isolated,
the inequality~\eqref{eq.diff} decomposes into $k$ independent inequalities, and each of them is consensus dichotomic. Hence any bounded solution of~\eqref{eq.diff} converges to a finite limit.

To prove the necessity, suppose that the inequality~\eqref{eq.diff} is dichotomic. We are going to show that if
$a_{ij}>0$, i.e. $j$ is connected to $i$ by an arc, then a walk from $i$ to $j$ exists (and hence $i$ and $j$ belong to the same component). Suppose the contrary and consider the set $S$ of all nodes, connected to $j$ by walks (including $j$ itself); by assumption $i\not\in S$.
We now consider an extension of Example~\ref{ex.trivial}. For $M>1$ being so large that $(M-1)a_{ij}\ge 2\sum_{k\not\in S}a_{ik}+1$, we
define the function $x(t)\in\r^N$ as follows:
\[
x_k(t)=
\begin{cases}
\sin t,\quad &k=i\\
M,\quad &k\in S\\
-1,\quad &k\not\in S\cup\{i\}.
\end{cases}
\]
We are going to show that $x(t)$ is a solution to~\eqref{eq.diff}. When $k\in S$, $a_{km}(x_m(t)-x_k(t))\equiv 0$
for any $m$ (indeed, if $m\not\in S$ then $a_{km}=0$, and otherwise $x_m(t)=x_k(t)=M$). Hence
$0=\dot x_k(t)=\sum_{m=1}^Na_{km}(x_m(t)-x_k(t))\,\forall k\in S$.
Obviously, $\dot x_k(t)=0\le \sum_{m=1}^Na_{km}(x_m(t)-x_k(t))$ for any $k\not S\cup\{i\}$. Finally, $\sum_{m\in S}a_{im}(x_m(t)-x_i(t))\ge (M-1)a_{ij}\ge -\sum_{m\not\in S}a_{im}(x_m(t)-x_i(t))+1$
and thus $\dot x_i(t)\le 1\le\sum_{m=1}^Na_{im}(x_m(t)-x_i(t))$. Hence the assumption $i\not\in S$ implies the existence of a non-converging bounded solution to~\eqref{eq.diff}, which is a contradiction. \epf

\subsection{Proof of Theorem~\ref{thm.3}}

For technical reasons, it is easier to prove the dichotomy and consensus dichotomy of the reversed inequality~\eqref{eq.diff1}. Introducing the ordering~\eqref{eq.ordering} of $x(t)$,~\eqref{eq.ordering-diff} implies that
\[
\dot y(t)\ge -L[B(t)]y(t),\quad b_{ij}(t)=a_{k^i(t)k^j(t)}(t).
\]
Retracing the proof of Theorem~1 in~\citep{TsiTsi:13} one can show that 1) any bounded solution $x(t)$ converges to a finite limit (as the vector $y(t)$ converges); 2) the functions
$\dot x_i$ and $a_{ij}(x_j-x_i)$ are $L_1$-summable for any $i,j$; 3) this implies consensus whenever the graph is ISC.
The necessity of ISC condition for the consensus dichotomy follows from Theorem~\ref{thm.2}.
\epf

\subsection{Proof of Theorem~\ref{thm.4}}

Thanks to the well-known result from~\cite[Theorem~1]{Moro:04}, any solution of~\eqref{eq.proto} reaches consensus with the ``leader'' $s$, that is, $x(t)\to x_s(0)\ones_N$, where the convergence is exponential and the convergence rate can be explicitly found. Introducing the Cauchy transition matrix of~\eqref{eq.proto} $\Phi(t,s)$, this implies that $\lim\limits_{t\to\infty}\Phi(t;0)$ is a matrix, whose
$s$th column equals $\ones_N$ and the other columns are zero. Consider a solution of~\eqref{eq.diff}, such that $x_i(s)\ge x_s(0)\,\forall i$, i.e. $x(t)\ge x_s(0)\ones_N$. Applying Corollary~\ref{cor.neg} to $f(t)\dfb \dot x(t)+L(t)x(t)\le 0$, we have
$$
x_s(0)\ones_N\le x(t)\le\Phi(t;0)x(0)\xrightarrow[t\to\infty]{}x_s(0)\ones_N.\qed
$$

\section{Conclusions}\label{sec.concl}

In this paper, we examine linear differential inequalities~\eqref{eq.diff}, arising in various problems of multi-agent coordination. An important property of such inequalities, established in this paper, is their \emph{consensus dichotomy}: under mild connectivity assumptions any bounded solution converges to a consensus equilibrium point. The dichotomy
criteria allow to analyze stability of many protocols for target aggregation, containment control, target surrounding and distributed optimization in a unified way. The results of this paper can be extended to discrete-time, or \emph{recurrent} inequalities $x(t+1)\le A(t)x(t)$, where $A(t)$ are row-stochastic matrices~\citep{ProCao2017-3}. Their extensions to the delayed inequalities and inequalities of second and higher orders are subject of ongoing research.


\bibliographystyle{elsarticle-harv}
\bibliography{consensus}

\end{document}